\documentclass[aip,sd,amsmath,amssymb,reprint]{revtex4-1}

\usepackage{amsmath}    
\usepackage{graphicx}   
\usepackage{verbatim}   
\usepackage{xcolor}      
\usepackage{subfigure}  
\usepackage{hyperref}   
\usepackage{diagbox}
\usepackage{subfigure}
\usepackage[font=footnotesize]{caption}
\draft

 \makeatletter
    \renewcommand\@make@capt@title[2]{%
     \@ifx@empty\float@link{\@firstofone}{\expandafter\href\expandafter{\float@link}}%
      {\textbf{#1}}\@caption@fignum@sep#2\quad}%
    \makeatother
   
\makeatletter
\renewcommand{\fnum@figure}{\textbf{Figure~\thefigure}}
\makeatother

\begin{document}

\title{Self-diffusion coefficient of the square-well fluid from molecular dynamics within the constant force approach}

\author{Alexis Torres-Carbajal}
\email[]{alpixels@fisica.ugto.mx}
\affiliation{Divisi\'on de Ciencias e Ingenier\'ias, Campus Le\'on, Universidad de Guanajuato, Loma del Bosque 103, Lomas del Campestre, 37150 Le\'on, Guanajuato, Mexico.}
\author{Victor M. Trejos}
\affiliation{Instituto de Qu\'imica, Universidad Nacional Aut\'onoma de M\'exico, Apdo. Postal 70213, Coyoac\'an 04510, Ciudad de M\'exico, M\'exico.} 
\author{Luz Adriana Nicasio-Collazo}
\affiliation{Divisi\'on de Ciencias e Ingenier\'ias, Campus Le\'on, Universidad de Guanajuato, Loma del Bosque 103, Lomas del Campestre, 37150 Le\'on, Guanajuato, Mexico.}
%

\date{\today}

\begin{abstract}
	We present a systematic study of the self-diffusion 
	coefficient for a fluid of particles 
	interacting via the square-well pair potential by means of molecular 
	dynamics simulations in the canonical $ (N, V, T) $ ensemble. The discrete 
	nature of the interaction potential is modeled through the 
	constant force approximation and the self-diffusion 
	coefficient is determined for several packing fractions 
	at super critical thermodynamic states. The 
	dependence of the self-diffusion coefficient with the potential 
	range $ \lambda $ is analyzed in the range of $ 1.1 \leq \lambda \leq 1.5 $. 
	The obtained molecular dynamics simulations results are in agreement with the 
	self-diffusion coefficient predicted with the Enskog method. Additionally, we show that 
	diffusion coefficient is very sensitive to the potential range, $ \lambda $,
	at low densities leading to a density dependence of this 
	coefficient not shared with other macroscopic properties such as 
	the	equation of state. The constant force approximation used in this 
	work to model the discrete pair potentials has shown to be an excellent scheme to compute transport properties using standard computer simulations. 
	Finally, the simulation results presented here 
	are resourceful to improving theoretical approaches, such as the Enskog method.  
\end{abstract}

\maketitle

\section{\label{SecI}Introduction}
The square-well (SW) pair potential has been widely used in statistical mechanics
in both theoretical approaches \cite{Barker1967,Smith1970,Smith1971,Henderson1976,Henderson1980,Carley1977,Carley1981,Carley1983,DelRio1983,DelRio1985,DelRio1987,DelRio1987v2,DelRio1989,DelRio1991,Gil1996} and computer simulations 
schemes \cite{Rotenberg1965,Lado1968,Rosenfeld1975,Scarfe1976,Vega1992}. 
This simple model has a repulsive and an attractive contribution, thus, the main feature of this potential is the 
ability to control, independently, the energy and range interaction between
molecules \cite{Paschinger2005}, this feature gives us the opportunity to characterize different systems of interest, ranging from simple liquids to complex fluids \cite{Alejandro1997} and even 
colloidal suspensions \cite{Duda2009}. Many of those fluids are being direct analogies of real substances. Thus, nowadays, the SW fluid is used to gain insight about the 
thermodynamics and phase behavior of ideal and real fluids \cite{Clare1999,Clare2010}.

However, results about transport properties of the SW pair potential 
are scarce, this lack of information is mainly due 
to the technical issues with dealing, in a dynamical way, with non-continuous 
potentials.
In general, the transport properties 
are of interest in academic and industrial areas, e.g., 
for the design or manipulation of processes where not only the initial 
and final state of the system are of interest, furthermore, a complete 
thermodynamic description requires detail information about how the 
system goes from a steady state to another one, i.e., how a 
perturbation acts over an equilibrium state. In particular, 
one of the most relevant transport properties to characterize the 
mass transfer is the diffusion coefficient. This coefficient is a 
macroscopic  measure of the particles 
tendency to drift through a system under the action of an external 
constant force. In the case of a system free of external fields, the drift of 
particles in the system is determined by choosing some particle inside 
the fluid and consider all other particles as the external field 
for the tagged particle. In this scenario, the diffusion is described 
by the so-called self-diffusion coefficient $ D $ \cite{Millat1996}. 

From a molecular point of view, the computation of $ D $ requires the knowledge of the particle positions (or velocities) at every time instant \cite{Allen1991}, thus, a theoretical description explicitly time dependent or one capable to give dynamical information based on static information is required. Nevertheless, dynamical information 
can be obtained directly from a computer simulation, in fact, the first attempt to compute $ D $ for a SW fluid was done by means of the so-called Event-Driven Molecular Dynamics (ED-MD) \cite{Alder1959,Krekelberg2007}. 
However, the ED-MD algorithm implementation is not straightforward. On the other hand, Molecular Dynamics (MD) simulations has been used to achieve,  with a high degree of numerical accuracy, the determination of different transport properties \cite{Dysthe1999,Meler2004I,Meler2004II}. The main advantage of MD  resides in the easy implementation, however, it is only suitable for continuous potentials, since it requires the knowledge of the force between particles, which is determined by the derivative of the interaction potential \cite{Allen1991,Frenkel}. 

Recently, Orea and Odriozola have proposed the so-called constant force approach \cite{Orea2013} 
, to deal with any non-continuous interaction potential. 
Padilla and Benavides have used this approach to compute the 
liquid-vapor phase diagram of the SW fluid only for a range interaction 
$ \lambda = 1.5 $ \cite{Padilla2017}. Their results shows a 
remarkably agreement with previous simulation results, see 
Ref. \cite{Padilla2017} and references therein.
Thus, in this endeavor we used MD simulations to determine the time 
evolution of the SW fluid particles. The discontinuities of the 
pair potential are treated by means of the constant force approach 
(CFA), i.e., we remove the discontinuities of the pair potential 
by approaching them with linear functions, whose derivative is a 
constant value \cite{Orea2013,Reyes2016,Padilla2017}. Details of the CFA and the SW parametrization are discussed further bellow. 
Our main interest resides in the determination of $ D $ for several packing 
fractions and different values of the range potential, $ \lambda $, for the SW 
fluid. The main purpose of this work is to  
provide MD simulations results of $ D $ for a wide range of fluid densities 
and several ranges of the interaction potential. Besides, the simulation 
results presented in this work are compared with the Enskog method for SW fluids.

This work is organized into five sections as follows.
In Sec. \ref{SecII} the self-diffusion coefficient and its 
implementation by using the Enskog method and MD simulation is presented.
Then, in Sec. \ref{SecIII}, the CFA in the context of the MD simulation 
technique is discussed. Self-diffusion coefficient results from MD and the corresponding comparison with the Enskog approach are reported in Sec. \ref{SecIV}. Finally, in Sec. \ref{SecV}, we offer some concluding remarks.

\section{\label{SecII}Self-diffusion coefficient}
As was mentioned in Sec. \ref{SecI},
the self-diffusion coefficient $ D $ is a measure 
of the molecules tendency to drift trough a system. For a system 
without the influence of external forces, one particle inside 
the fluid is moved by the force that experiences due to the 
mass gradient at local scale \cite{Millat1996}. 
From the theoretical point of view, the Enskog theory have been widely 
used to determine transport 
properties as the self-diffusion coefficient of dense 
hard-sphere (HS) fluids \cite{Dymond1985,Speedy1987}. 
Modifications of such approach 
have raised to deal with the SW fluid \cite{Higgins1958,Davis1961,McLaughlin1966,Brown1971}. 
Nevertheless, one of the most used expressions \cite{XinYu2001} to 
determine $ D $ was proposed by Davis {\it et.al.}\cite{Davis1961,McLaughlin1966} 
and express the self-diffusion coefficient as
\begin{equation}
	D = \frac{3}{8 \rho \sigma^{2}} \left( \frac{k_{B}T}{\pi m}\right)^{1/2} \left[ g \left(\sigma\right) - \lambda^{2} g\left(\lambda \sigma\right) \Xi \right]^{-1}, 
	\label{eq: 1}
\end{equation}   
where $ \rho $ is the density number of the fluid, $ m $ is the mass of one particle, $ k_{B} $ is the Boltzmann constant, $ T $ is the absolute 
temperature, $ g\left(\sigma\right) $ and $ g\left(\lambda \sigma\right) $ are the equilibrium radial distribution function (RDF) evaluated at the contact value
$ \sigma $ and discontinuity value of $ \lambda \sigma $, respectively. 
For a SW fluid, $ \Xi $ is given by,
\begin{equation}
	\Xi = \exp \left( \frac{\epsilon }{k_{B}T}\right) - \frac{\epsilon}{2 k_{B}T} - 2J,
	\label{eq: 2}
\end{equation}
where $ \epsilon $ is the depth of the SW potential and $ J $ is given by
\begin{equation}
	J = \int\limits_{0}^{\infty} y^{2} \left(y^{2} + \frac{\epsilon}{k_{B}T}\right)^{1/2} \exp \left( -y^{2}\right) ~ dy.
	\label{eq: 3}
\end{equation}

As one can see, Eq. \eqref{eq: 1} is not an explicitly function of the time, 
it only needs static information given trough the RDF is needed. 
For SW fluids the contact and the 
discontinuity value of RDF can be obtained by means of analytical 
expressions as the first order perturbation term of the pressure 
equation \cite{Henderson1976,Chang1994I,Chang1994II} and from Monte Carlo 
simulation results \cite{Barker1971}. However, as 
Duffy {\it et. al.}\cite{Duffy1991} pointed out, Eq. \eqref{eq: 1} it 
is not enough to describe the self-diffusion coefficient at low and high 
fluid densities. The approximation given by Eq. \eqref{eq: 1} was 
corrected by Yu {\it et. al.} \cite{XinYu2001} using MD simulation results \cite{Alley1975,Michels1982,Michels1975}
and the self-diffusion coefficient is rewritten as,
\begin{equation}
D = \frac{3}{8 \rho \sigma^{2}} \left( \frac{k_{B}T}{\pi m}\right)^{1/2} \left[ g \left(\sigma\right) f_{R} \left( \rho^{*} \right) - \lambda^{2} g\left(\lambda \sigma\right) \Xi f_{S} \left( \rho^{*} \right) \right]^{-1},
\label{eq: 4}	 
\end{equation}     
where $ f_{R} \left( \rho^{*} \right) $ is the correction function to 
compute accurate results of the HS fluid in the moderate and high-density 
ranges. Also, by using MD results, Speedy \cite{Speedy1987} proposed 
the following correction function
\begin{equation}
f_{R} \left( \rho^{*} \right) = \frac{\left( 1 - \eta \right)^{3}}{\left( 1 - 0.5 \eta \right) \left( 1 - \rho^{*}/1.09 \right) \left( 1 + 0.4 \rho^{*~2} - 0.83 \rho^{*~4} \right)}.
\label{eq: 5}	
\end{equation} 

On the other hand, the correct temperature dependence of the $ D $ is improved 
with the correction function $ f_{S} \left( \rho^{*} \right) $, also determined by means of MD simulation results \cite{Alley1975,Michels1982,Michels1975}, given by
\begin{equation}
f_{S} \left( \rho^{*} \right) = 70.771 \rho^{*~3} - 58.971 \rho^{*~2} + 19.903 \rho^{*} - 1.3708,
\label{eq: 6}	
\end{equation}
where, $ \eta $ stands for the packing fraction and the reduced density $ \rho^{*} $ is given by $ \rho^{*} = 6 \eta / \pi $. 
In this work, we use the Eq. \eqref{eq: 4} 
as the theoretical value of $ D $, and the predictions given by this approximation 
are compared with the simulation results. 

It is worth to point out that there are others approaches for  $ D $ based 
on theoretical or empirical formulations that reproduces the 
computer simulation data for particular values of the 
interaction potential parameters and some thermodynamic states \cite{Liua1998}. However, 
those formulations of $ D $ has large deviations even between them \cite{XinYu2001}, this 
issue is due to the use of one particular approach to predict results of 
$ D $ far away of the thermodynamic states and potential parameters used to 
calibrate it. In this sense, such approximations are limited to some thermodynamic states or certain values of  parameters in the pair potential. Thus, the election of an equation for $ D $ must to take into account details 
	about its deduction. 
Of course, this is not an easy task, furthermore, it is a limitation for a systematic study of the self-diffusion coefficient of SW fluids. 

Nevertheless, if one is able to compute and follow the time evolution, of any, the  position or the velocity of particles that compose the SW fluid, the self-diffusion coefficient can be computed without any approximation or assumption. Thus, the determination of $ D $ requires explicitly knowledge 
of the system time evolution, in this sense, such coefficient can be computed 
by two different, but analogue, routes. One of them are the Green-Kubo relation \cite{Zwanzig1965}, which is defined as a time integration of the velocity auto-correlation function (VACF) \cite{Allen1991}, 
\begin{equation}
D = \frac{1}{d} \int\limits_{0}^{\infty} \langle {\bf v}\left( t \right)
{\bf v}\left( 0 \right) \rangle ~ dt,
\label{eq: 7} 
\end{equation}  
where $ \langle {\bf v}\left( t \right)
{\bf v}\left( 0 \right) \rangle $ is the VACF and $ d $ is the system  
dimensionality. The second route is the Einstein 
relation \cite{Allen1991}, that uses the mean square displacement (MSD) to determine $ D $, and it is given by,
\begin{equation}
2dD = \lim\limits_{t \rightarrow \infty} \frac{W\left(t\right)}{t},
\label{eq: 8}
\end{equation}
where $ W\left( t \right) \equiv \langle \left[ {\bf r}\left( t \right) - {\bf r}\left( 0 \right) \right]^{2} \rangle $ is the MSD. Nowadays, for a continuous potential the computation of VACF or MSD is an 
easy task by means of standard MD simulations that also has 
the advantage to control, either, the temperature or pressure of the fluid
\cite{Allen1991,Frenkel}, however, if the interaction potential has a 
non-continuous shape, as the SW potential, see Fig. \ref{Fig: 1}, the 
MD technique can not be used, since it requires the knowledge of the force 
between molecules, that it is not-well defined at the discontinuities of 
the interaction potential. 

In this work, we use the Eq. \eqref{eq: 8} and MD simulations within the 
CFA to, i) carried out the time evolution 
of the SW particles to compute the MSD and ii) to determine $ D $ 
with the long time behavior of MSD in a systematic way, for several 
packing fractions and different values of $ \lambda $.

\section{\label{SecIII}Constant force approach and computer simulation.}
In order to compute $ D $ for SW fluids 
we use MD simulations in the canonical $ (N,V,T) $ ensemble. The Hamiltonian of the system is given by,
\begin{equation}
H = \frac{1}{2m} \sum_{i=1}^{N} {\bf p}_{i}^{2} + \sum_{i=1; i \neq j}^{N} \mathcal{U} \left( \vert {\bf r}_{i} - {\bf r}_{j} \vert \right),
\label{eq: 9}
\end{equation}
where $ {\bf p}_{i} $ and $ {\bf r}_{i} $ are the linear momentum and position,
respectively, of the $ i $-th fluid particle. We define $ r_{ij} \equiv \vert {\bf r}_{i} - {\bf r}_{j} \vert $ as the separation distance between the center of mass of particles $ i $ and $ j $. The term $ \mathcal{U} $ in Eq. \eqref{eq: 9} is the interaction pair potential that in this contribution is taken 
as the SW potential given by,
\begin{equation}
\mathcal{U} \left( r \right) = \left\lbrace 
\begin{array}{l}
\infty \hspace*{1cm} r \leq \sigma \\
-\epsilon \hspace*{0.9cm} \sigma < r < \lambda \sigma \\
0 \hspace*{1.2cm} r \geq \lambda \sigma,
\end{array}
\right.
\label{eq: 10}
\end{equation}
where $ \sigma $ is the diameter of the hard-core, $ \epsilon $ is  the 
magnitude of the attractive part of the potential and $ \lambda \sigma $ 
is the diameter of the surrounding well. In the framework of the CFA the SW pair potential, see Eq. \eqref{eq: 10}, is 
parametrized as
\begin{equation}
\mathcal{U}\left( r_{ij} \right) = \mathcal{U}_{HS}\left( r_{ij}; \sigma, \lambda_{o}, \epsilon_{1} \right) + \mathcal{U}_{SW}^{D}\left( r_{ij}; \sigma, \lambda_{1}, \epsilon_{2} \right),
\label{eq: 11}
\end{equation} 
with $ \mathcal{U}_{HS} $ being the hard-sphere potential and $ \mathcal{U}_{SW}^{D} $ is the discrete step of the square-well interaction, respectively. Explicitly,  such contributions are given by
\begin{equation}
\mathcal{U}_{HS}\left( r_{ij}; \sigma, \lambda_{o}, \epsilon_{1} \right) = - \alpha_{o} \left( r_{ij} - \lambda_{o} \sigma \right) + \epsilon_{1}, \hspace*{0.6cm} r \leq \sigma,
\label{eq: 12}
\end{equation}
where $ \alpha_{o} $ is the slope stiffness with value, in reduced units, of $ \alpha_{o}=1000=\alpha_{1} $, and for this contribution $ \epsilon_{1} = 0$. On the other hand, the contribution $ \mathcal{U}_{SW}^{D} $ within the framework of the CFA is given by,
\begin{equation}
\mathcal{U}_{SW}^{D}\left( r_{ij} \right) = \left\lbrace 
\begin{array}{l}
\epsilon_{2}, \hspace*{3.2cm} \lambda r_{o} < r_{ij} < \lambda_{1} \sigma, \\
B_{i}\alpha_{i} \left(r_{ij} - \lambda_{1}\sigma \right) + \epsilon_{2}, \hspace*{0.3cm} \lambda_{1}\sigma \leq r_{ij} \leq \lambda r_{1},
\end{array}
\right.
\label{eq: 13}
\end{equation}
where $ \lambda r_{o} = \lambda_{o}\sigma + \delta_{o} $, $ \lambda r_{1} = \lambda_{1} \sigma + \delta_{1} $, $ \epsilon_{2}=-1 $, $ \delta_{o} = 0 $ and $ \delta_{1} = \left(\epsilon_{2} - \epsilon_{1}\right)/B_{i}\alpha_{i} $.  $ B_{i} $ gives the sign of the slope $ \alpha_{i} $ for the linear function at each discontinuity, and it is defined as
\begin{equation}
B_{i} = \left\lbrace
\begin{array}{l}
1 \hspace*{1.3cm} \text{if } \epsilon_{i+1} > \epsilon_{i} \\
-1 \hspace*{1cm} \text{if } \epsilon_{i+1} < \epsilon_{i},
\end{array}
\right.
\label{eq: 14}
\end{equation}

\begin{figure}[htp!]
	\includegraphics[scale=0.5]{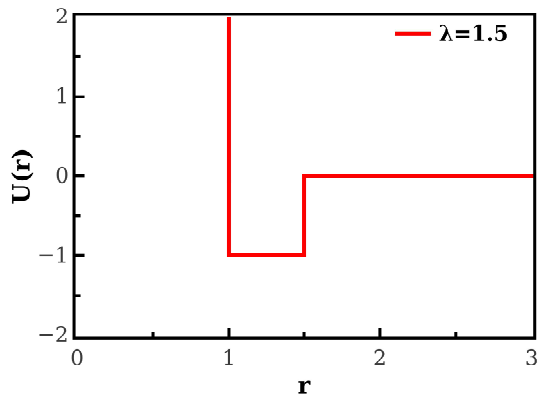}
	\caption{Schematic representation of the SW pair potential	
		 of range $ \lambda=1.5 $. In the framework of the CFA the potential is approximated with $ \alpha_{o}= \alpha_{1} = 1000 $, $ \epsilon_{1}=0 $ and $ \epsilon_{2}=-1 $, respectively.}
	\label{Fig: 1}
\end{figure}

Thus, the SW potential representation in the CFA framework can be seen in Fig. \ref{Fig: 1}. 
We stress the fact that the derivative of the potential at the discontinuities has a constant value given by $ \frac{d}{d r} \mathcal{U}_{SW}^{D} \left( r \right)\arrowvert _{\lambda_{i} \sigma} = B_{i} \alpha_{i} $, hence, the force at these points is constant as well. 

The studied system is composed by $ N = 1372 $ spherical particles, initially distributed on a FCC configuration with random velocities that satisfy the equipartition theorem \cite{Allen1991}. We use standard reduced units of length, temperature and time defined as $ r^{*} \equiv r/\sigma $, $ T^{*} = k_{B}T/ \epsilon $ and $ t^{*} = t/\tau $, respectively. Where, $ \sigma, m, \epsilon $ and $ \tau = \sqrt{m \sigma^{2} / \epsilon} $ are the usual units of length, mass, energy and time, respectively. In the same line, the packing fraction is defined as $ \eta = N \pi \sigma^{3} / 6 V $, with $ V $ the simulation box volume. Thus, the reduced density is given by $ \rho^{*} = 6 \eta / \pi $.

The equations of motion are integrated with the velocity Verlet algorithm \cite{Swope1982} employing a time step $ \Delta t = 1 \times 10^{-5} \tau $ to guarantee the numerical stability of the CFA. It is worth to point out that in difference with previous works \cite{Orea2013,Padilla2017},SecIV we do not use an external input table to compute any the interaction potential or the force between the fluid particles, instead such contributions are explicitly computed at each time step in order to avoid numerical inaccuracies. In all cases, we performed $ 1.1 \times 10^{9} $ integration steps, where $ 1 \times 10^{8} $ integration steps were carried out to reach thermal equilibrium and the subsequent time steps were used to compute the static and dynamic properties like the radial distribution function and the mean-squared displacement, respectively. The statical uncertainties associated with the time correlation function are obtained according to the procedure describe in Ref. \cite{Swope1982}. The system temperature is kept fixed by a simple velocity scaling as the thermostat \cite{Frenkel}.    

\section{\label{SecIV}Self-diffusion coefficient of the square-well fluid.}
\subsection{\label{SecSubI}Mean-square displacement}
\begin{figure}
	\includegraphics[scale=0.48]{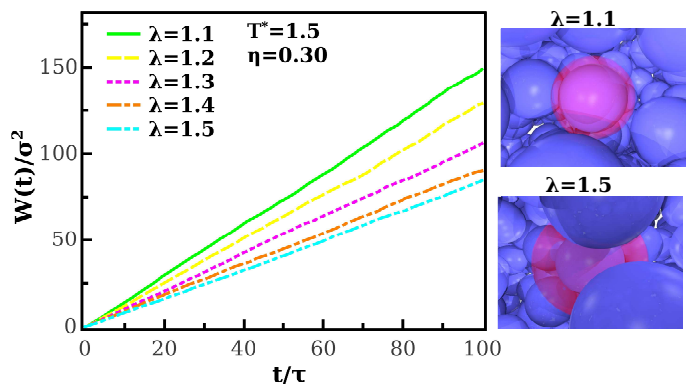}
	\caption{Mean square displacement for SW fluid particles 
		at several values of the range interaction potential 
		$ \lambda = 1.1, 1.2, 1.3, 1.4, 1.5$. In all cases the fluid is at the temperature $ T^{*} = 1.5 $ and packing fraction $ \eta = 0.30 $. The snapshot presented correspond to range interaction (top) $ \lambda =1.1 $ and (bottom) $ \lambda =1.5 $, respectively, at the thermodynamic conditions mentioned.}
	\label{Fig: 2}
\end{figure}

As was mentioned in Sec. \ref{SecII}, in order to determine the self-diffusion 
coefficient by means of the Eq. \eqref{eq: 8}, the knowledge of the MSD is 
needed, which is computed directly from the MD simulations. The behavior of MSD 
for several values of $ \lambda $ at the reduced temperature $ T^{*}=1.5 $ 
and the packing fraction $ \eta = 0.3 $ is shown in Fig. \ref{Fig: 2}. 

As one can see in Fig. \ref{Fig: 2}, the magnitude of the MSD decreases 
as the attraction range of the SW potential increases, those differences are 
magnified at low values of $ \lambda $. Since the MSD can be understood as the 
measure of one particle position deviation with respect to a reference 
position over time, in this scenario, the decrease of the MSD implies that 
fluid particles are more localized. This behavior is due to an increment 
of $ \lambda $, or equivalent, given any particle, an increment of the 
surrounding fluid particles. This observation can be corroborated with 
the snapshots in Fig. \ref{Fig: 2}, where we have selected a random particle, 
which is surrounding  by an attractive spherical (red) shell of diameter 
$ \lambda \sigma $. As can be seen in the snapshots of this figure, as the 
$ \lambda \sigma $ value increase the number of neighbor particles 
also does for a system at the same thermodynamic state. Of course, 
for a fixed values of $ \lambda \sigma $ and $ T^{*} $ the MSD has also 
a magnitude decrement as the packing fraction (density) increases. 

\subsection{\label{SecSubII}Self-Diffusion coefficient}
\begin{figure}
	\includegraphics[scale=0.55]{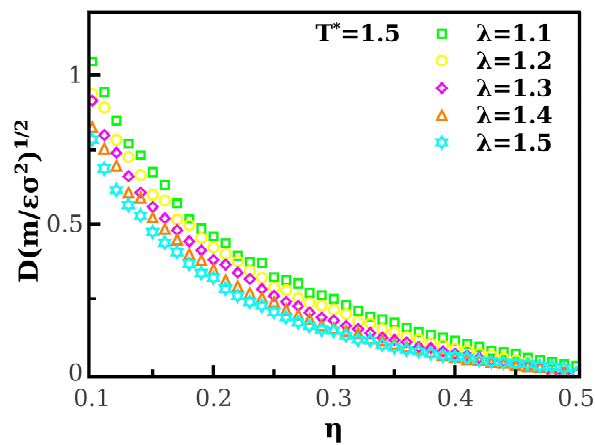}
	\caption{Self-diffusion coefficient $ D $ for SW fluids as a
		function of the packing fraction at different values of the interaction 
		potential range $ \lambda = 1.1, 1.2, 1.3, 1.4, 1.5$. In all cases, the reduced temperature is $ T^{*}=1.5 $.}
	\label{Fig: 3}
\end{figure}

From the long time behavior of the MSD we extract the value of $ D $ 
as a function of the packing fraction for different values of $ \lambda $, 
see Fig. \ref{Fig: 3}. Our mathematical approach to determine 
$ D $ lead us an estimation of it with a high degree of numerical accuracy, 
that gives us, for all cases, error bars much more smaller than the 
symbol size. 

As one could expect, $ D $ decreases as the packing fraction of the fluid 
increases independently of the potential interaction range. However and 
surprisingly, in general, the magnitude of $ D $ is greater for low values 
of $ \lambda $ if one sees at the same value of the packing fraction. 
These differences decreases as the attractive interaction range increases 
and as it is expected such differences are lost at the high concentration 
regime where the fluid is dynamical arrested. The first fluid that 
experiences the arrest is the one with the greater attraction range 
of this study, $ \lambda = 1.5 $, and it happens, approximately, at $ \eta \approx 0.49 $.   

For the self-diffusion coefficient, as can one can see in Fig. \ref{Fig: 3}, 
the interaction potential parameters of the SW fluid, $ \lambda $ and $ \epsilon $, are highly relevant 
at low and intermediate packing fractions, i.e., for $ \eta < 0.3 $. 
In the high concentration regime the hard-core interaction between 
particles dominates and the dynamical behavior tends to be similar, 
independently from the potential interaction range.  

\subsection{\label{SecSubIII}Test of the Enskog method for SW fluids}
\begin{figure}
	\includegraphics[scale=0.55]{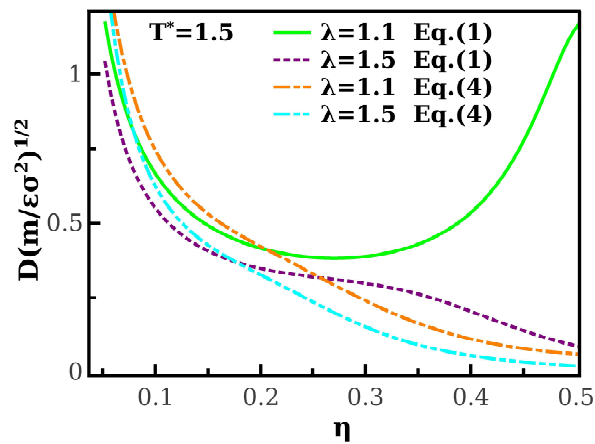}
	\caption{Comparison between the self-diffusion coefficient, $ D $, predicted 
		with  Eqs. \eqref{eq: 1} and \eqref{eq: 4} obtained with the Enskog method as a function of the packing fraction.}
	\label{Fig: 4}
\end{figure}

In Section \ref{SecII}, we summarized the expressions needed to compute 
$ D $ in the framework of the Enskog theory. Although, from theoretical 
point of view, there are several approaches to compute $ D $, the Eq. 
\eqref{eq: 1} is the first proposal given in terms of the Enskog theory, 
but, as has been demonstrated previously \cite{XinYu2001}, this 
expression it is not suitable to predict $ D $ at moderate and high 
concentrations. In fact, as one can see in Fig. \ref{Fig: 4}, the Eq. \eqref{eq: 1} 
totally fails at moderate concentration, i.e., $ \eta > 0.2 $, and low values 
of $ \lambda \sigma $, whereas as the attractive range interaction 
is increased an oscillatory behavior can be glimpsed at moderate 
concentrations. On the other hand, the predictions of Eq. \eqref{eq: 4} 
for $ D $ overcomes the issues aforementioned despite of at low 
concentrations predicts greater values of $ D $ than the Eq. \eqref{eq: 1}. 

\begin{figure}
	\includegraphics[scale=0.55]{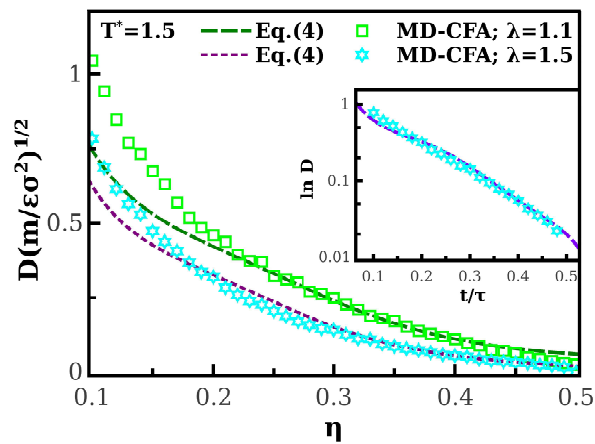}
	\caption{Comparison between the self-diffusion coefficient, $ D $, predicted with (lines) the Enskog method Eq. \eqref{eq: 4} and results from (dots) MD  within the CFA approach as a function of the packing fraction for the limit cases of our study $ \lambda = 1.1 $ and $ \lambda = 1.5 $ for the reduced temperature $ T^{*}=1.5 $.}
	\label{Fig: 5}
\end{figure}

Notwithstanding there exists different 
functional forms to predict $ D $ for the SW fluid, it is a matter 
of fact that the agreement between them is questioned. In this line of thoughts, the success of some particular approach depends on the 
assumptions done to its deduction. In this way, computer simulations 
are used to test such theoretical proposals, nevertheless, 
and despite of the plethora of results about SW fluid, its dynamical 
properties are far less studied. In Fig. \ref{Fig: 5}, we have shown a 
systematically study of self-diffusion coefficient and our computer simulation results 
are compared with those provided by Eq. \eqref{eq: 4}.    

In Fig. \ref{Fig: 5}, we compare the lowest 
and highest values of $ \lambda $ used in this work. We found a well 
qualitative agreement between the simulation results and the predictions 
given by Eq. \eqref{eq: 4}, see the inset of Fig. \ref{Fig: 5}. However, 
a closer inspection revels us differences between both results at low 
concentrations that are bigger than the acceptable tolerance. The 
Eq. \eqref{eq: 4} given by Yu \cite{XinYu2001} uses simulation 
results of the SW fluid at moderate and high concentrations to proposed 
the correction functions $ f_{R} \left( \rho^{*}\right) $ and $ f_{S} \left( \rho^{*}\right) $, given by Eqs. \eqref{eq: 5} and \eqref{eq: 6},  respectively. For that, one could not expect a good performance of this approach 
at low packing fraction. Thus, our simulation results can be used to calibrate theoretical approaches 
in order to properly describe the low concentration regime, 
however, for the time being, such analysis it is out of the scope of this contribution.

\section{\label{SecV}Concluding remarks}
In this paper, we have performed a systematically study of the 
self-diffusion coefficient, $ D $, 
of the SW fluid by means of molecular dynamics simulations within the 
constant force approach in the canonical $ (N, V, T) $ ensemble. From the 
analysis of the simulation data, we show that, 
for a fixed temperature and low density, the magnitude of $ D $ is 
higher for low values of the attractive interaction range 
than for the higher values of $ \lambda $, at the same thermodynamic 
states. The increment of the attractive range in the potential causes 
a higher localization of the fluid particles (see the snapshots in Fig. \ref{Fig: 2}), 
leading to a decrease of $ D $ independently of the packing fraction. On the other hand, our results 
exhibits a very well agreement with the theoretical predictions of the 
Enskog method corrected by Yu \cite{XinYu2001}  at 
moderated and high packing fractions, however, at low densities 
such agreement is lost. 

Furthermore, we have provided new simulation data results
for $ D $ in a wide range of densities 
and different values of the attractive range interaction not reported 
previously. Thus, this results can be used to improve the correction 
functions in the theoretical descriptions, or to provide new and more sophisticated empirical relations. As well, the CFA manifest itself as a valuable tool, not only 
to determine the statistical properties of non-continuous potential 
but also its dynamical behavior. Finally, the 
mathematical framework reported in this work can be employed for 
both the determination of structural and dynamical properties of 
liquids or even colloidal systems that are characterized with the SW potential \cite{Valadez2012,Zhao2017}.

\begin{acknowledgments}
	A. Torres-Carbajal thank the valuable discussions with A. Padilla (UG)
	about the CFA details. 
	V. M. Trejos also thank S. F.-Gerstenmaier (UG) for high performance 
	computer facilities (Fondos Mixtos CONACYT-CONCYTEG
	2011, PROMEP 2010) that have partially contributed to the 
	research results reported within this paper.
\end{acknowledgments}

\bibliographystyle{ieeetr}
\bibliography{Square.bib}

\end{document}